\documentclass[aps,prl,superscriptaddress,twocolumn,nofootinbib,floatfix]{revtex4} 
\usepackage{graphicx}
\begin{document}

\title{FIRST EVIDENCE OF NEW PHYSICS IN $\mathbf{b \leftrightarrow s}$ TRANSITIONS
\begin{figure}[htb!]
 \begin{center}
 \includegraphics[width=0.1\textwidth]{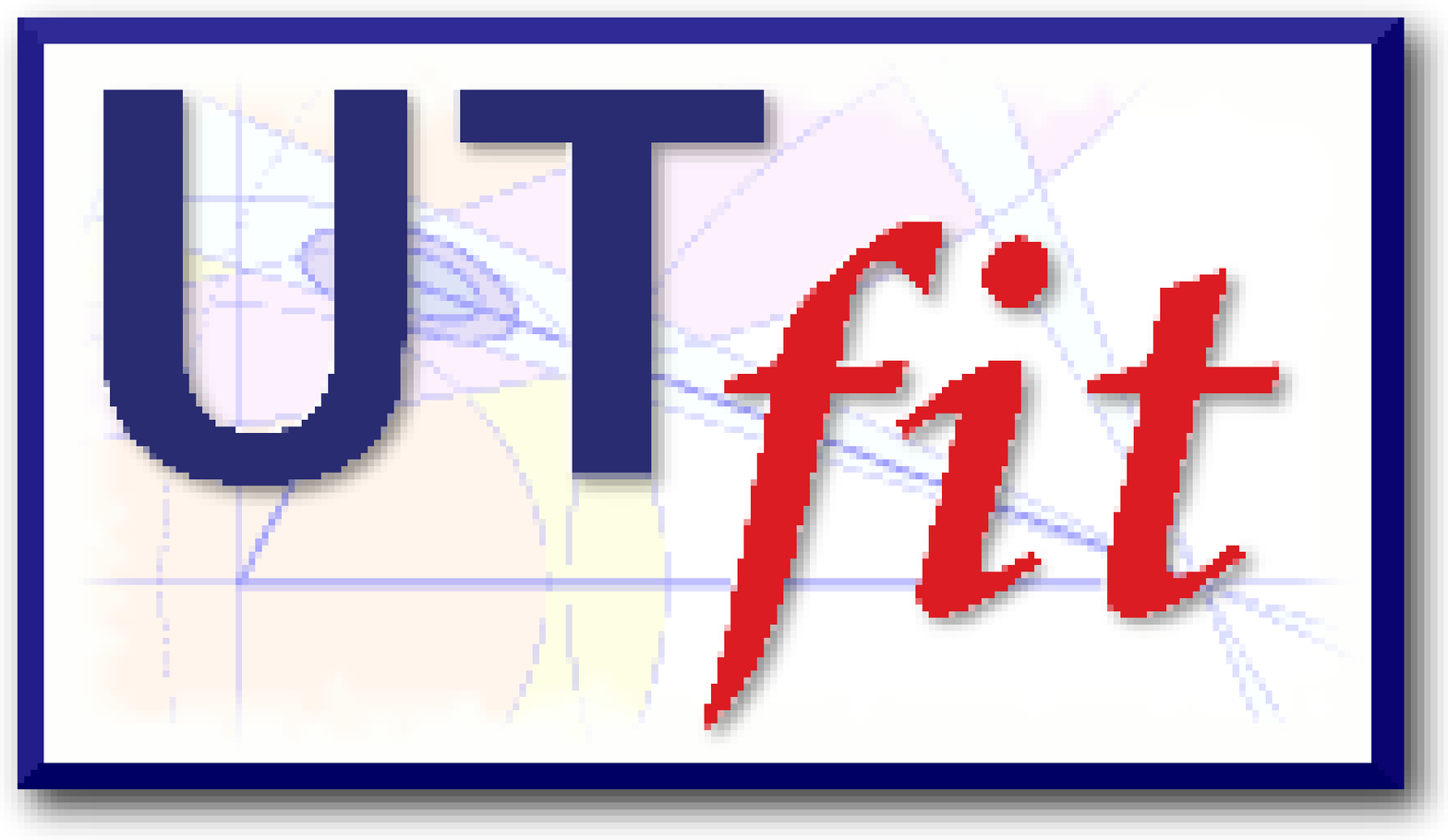}
 \end{center}
\end{figure}
\vspace*{-0.8cm}
}

\collaboration{\textbf{UT}\textit{fit} Collaboration}
\noaffiliation
\author{M.~Bona}
\affiliation{CERN, CH-1211 Geneva 23, Switzerland}
\author{M.~Ciuchini}
\affiliation{INFN,  Sezione di Roma Tre, I-00146 Roma, Italy}
\author{E.~Franco}
\affiliation{INFN, Sezione di Roma, I-00185 Roma, Italy}
\author{V.~Lubicz}
\affiliation{INFN,  Sezione di Roma Tre, I-00146 Roma, Italy}
\affiliation{Dipartimento di Fisica, Universit{\`a} di Roma Tre, I-00146 Roma, Italy}
\author{G.~Martinelli}
\affiliation{INFN, Sezione di Roma, I-00185 Roma, Italy}
\affiliation{Dipartimento di Fisica, Universit\`a di Roma ``La Sapienza'', I-00185 Roma, Italy}
\author{F.~Parodi}
\affiliation{ Dipartimento di Fisica, Universit\`a di Genova and INFN, I-16146
  Genova, Italy} 
\author{M.~Pierini}
\affiliation{CERN, CH-1211 Geneva 23, Switzerland}
\author{C.~Schiavi}
\affiliation{ Dipartimento di Fisica, Universit\`a di Genova and INFN, I-16146
  Genova, Italy} 
\author{L.~Silvestrini}
\affiliation{INFN, Sezione di Roma, I-00185 Roma, Italy}
\author{V.~Sordini}
\affiliation{Laboratoire de l'Acc\'el\'erateur Lin\'eaire, IN2P3-CNRS et
  Universit\'e de Paris-Sud, BP 34,
      F-91898 Orsay Cedex, France}
\author{A.~Stocchi}
\affiliation{Laboratoire de l'Acc\'el\'erateur Lin\'eaire, IN2P3-CNRS et
  Universit\'e de Paris-Sud, BP 34, 
      F-91898 Orsay Cedex, France}
\author{V.~Vagnoni}
\affiliation{INFN, Sezione di Bologna,  I-40126 Bologna, Italy}

\begin{abstract}
  We combine all the available experimental information on $B_s$
  mixing, including the very recent tagged analyses of $B_s \to J/\Psi
  \phi$ by the CDF and D{\O} collaborations. We find that the phase of
  the $B_s$ mixing amplitude deviates more than $3\sigma$ from the
  Standard Model prediction. While no single measurement has a
  $3\sigma$ significance yet, all the constraints show a remarkable
  agreement with the combined result.  This is a first evidence of
  physics beyond the Standard Model. This result disfavours New
  Physics models with Minimal Flavour Violation with the same
  significance.
\end{abstract}

\maketitle

In the Standard Model (SM), all flavour and CP violating phenomena in
weak decays are described in terms of quark masses and the four
independent parameters in the Cabibbo-Kobayashi-Maskawa (CKM)
matrix~\cite{CKM}. In particular, there is only one source of CP
violation, which is connected to the area of the Unitarity Triangle
(UT). A peculiar prediction of the SM, due to the hierarchy among CKM
matrix elements, is that CP violation in $B_s$ mixing should be
tiny. This property is also valid in models of Minimal Flavour
Violation (MFV)~\cite{mfv}, where flavour and CP violation are still
governed by the CKM matrix. Therefore, the experimental observation of
sizable CP violation in $B_s$ mixing is a clear (and clean) signal of
New Physics (NP) and a violation of the MFV paradigm. In the past
decade, $B$ factories have collected an impressive amount of data on
$B_d$ flavour- and CP-violating processes. The CKM paradigm has passed
unscathed all the tests performed at the $B$ factories down to an
accuracy just below $10\%$~\cite{utfit,ckmfit}.  This has been often
considered as an indication pointing to the MFV hypothesis, which has
received considerable attention in recent years.  The only possible
hint of non-MFV NP is found in the penguin-dominated $b\to s$
non-leptonic decays. Indeed, in the SM, the $S_{q\bar qs}$ coefficient
of the time-dependent CP asymmetry in these channels is equal to the
$S_{c\bar cs}$ measured with $b\to c\bar c s$ decays, up to hadronic
uncertainties related to subleading terms in the decay amplitudes.
Present data show a systematic, although not statistically
significant, downward shift of $S_{q\bar qs}$ with respect to
$S_{c\bar cs}$~\cite{bspeng}, while hadronic models predict a shift in
the opposite direction in many cases~\cite{deltas,review}.

From the theoretical point of view, the hierarchical structure of
quark masses and mixing angles of the SM calls for an explanation in
terms of flavour symmetries or of other dynamical mechanisms, such as,
for example, fermion localization in models with extra dimensions. All
such explanations depart from the MFV paradigm, and generically cause
deviations from the SM in flavour violating processes. Models with
localized fermions~\cite{localized}, and more generally models of
Next-to-Minimal Flavour Violation~\cite{NMFV}, tend to produce too
large effects in $\varepsilon_K$~\cite{df2gen,sacha}. On the contrary,
flavour models based on nonabelian flavour symmetries, such as $U(2)$
or $SU(3)$, typically suppress NP contributions to $s \leftrightarrow
d$ and possibly also to $b \leftrightarrow d$ transitions, but easily
produce large NP contributions to $b \leftrightarrow s$
processes. This is due to the large flavour symmetry breaking caused
by the top quark Yukawa coupling. Thus, if (nonabelian) flavour
symmetry models are relevant for the solution of the SM flavour
problem, one expects on general grounds NP contributions to $b
\leftrightarrow s$ transitions. On the other hand, in the context of
Grand Unified Theories (GUTs), there is a connection between leptonic
and hadronic flavour violation. In particular, in a broad class of
GUTs, the large mixing angle observed in neutrino oscillations
corresponds to large NP contributions to $b \leftrightarrow s$
transitions~\cite{guts}.

In this Letter, we show that present data give evidence of a $B_s$
mixing phase much larger than expected in the SM, with a significance
of more than $3\sigma$. This result is obtained by combining all
available experimental information with the method used by our
collaboration for UT analyses and described in Ref.~\cite{utfitbasic}.

We perform a model-independent analysis of NP contributions to $B_s$
mixing using the following parametrization~\cite{cfactors}:
\begin{eqnarray} C_{B_s}
  \, e^{2 i \phi_{B_s}} &=&\frac{A^\mathrm{SM}_s e^{-2 i \beta_s} +
    A^\mathrm{NP}_s e^{2 i (\phi^\mathrm{NP}_s - \beta_s)}}{A^\mathrm{SM}_s
    e^{-2 i \beta_s}}= \nonumber \\
  &=& \frac{\langle
    B_s|H_\mathrm{eff}^\mathrm{full}|\bar{B}_s\rangle} {\langle
    B_s|H_\mathrm{eff}^\mathrm{SM}|\bar{B}_s\rangle}\,, \quad 
  \label{eq:paranp}
\end{eqnarray}
where $H_\mathrm{eff}^\mathrm{full}$ is the effective Hamiltonian
generated by both SM and NP, while $ H_\mathrm{eff}^\mathrm{SM}$ only
contains SM contributions. The angle $\beta_s$ is defined as $\beta_s=
\arg(-(V_{ts}V_{tb}^*)/(V_{cs}V_{cb}^*))$ and it equals $0.018 \pm
0.001$ in the SM.\footnote{We are using the usual CKM phase convention
  in which $V_{cs}V_{cb}^*$ is real to a very good approximation.}

\begin{table}[h]
\begin{center}
\begin{tabular}{lcc}
\hline
$\Delta m_s$ [ps$^{-1}$]                & 17.77 $\pm$ 0.12  &  \cite{dmsCDF}\\
$A_\mathrm{SL}^s\times 10^2$            &  2.45 $\pm$ 1.96  &   \cite{ASLD0}\\
$A_\mathrm{SL}^{\mu\mu}\times 10^3$     & -4.3  $\pm$  3.0  & 
\cite{ACHD0,ASLCDF}  \\
$\tau_{B_s}^\mathrm{FS}$ [ps]           & 1.461 $\pm$ 0.032 &
\cite{tauBsflavspec}\\
\hline
$\phi_s$                          &  \texttt{http://tinyurl.com/2f9rtl}  &
 \cite{CDFTAGGED} \\
$\Delta\Gamma_s$            &  \texttt{http://tinyurl.com/2f9rtl}  &
 \cite{CDFTAGGED}\\
\hline
$\phi_s$ [rad]                          & 0.60 $\pm$ 0.27   & \cite{D0TAGGED}\\
$\Delta\Gamma_s$ [ps$^{-1}$]            & 0.19 $\pm$ 0.07   & \cite{D0TAGGED}\\
$\tau_{B_s}$ [ps]                       & 1.52 $\pm$ 0.06   & \cite{D0TAGGED}\\
\multicolumn{3}{l}{$C_{\phi_s, \Delta\Gamma_s} = -0.042$ \quad  
$C_{\phi_s, \tau_{B_s}} = -0.571$ \quad 
$C_{\tau_{B_s}, \Delta\Gamma_s} = 0.23$} \\
\hline
\end{tabular}
\end{center}
\caption {Input parameters used in the analysis. We also show the
correlation coefficients $C$s of the measurements of $\phi_s$, $\Delta\Gamma_s$
and $\tau_{B_s}$ from ref.~\cite{D0TAGGED}.}
\label{tab:input}
\end{table}

\begin{figure}[ht]
\begin{center}
\includegraphics[width=0.23\textwidth]{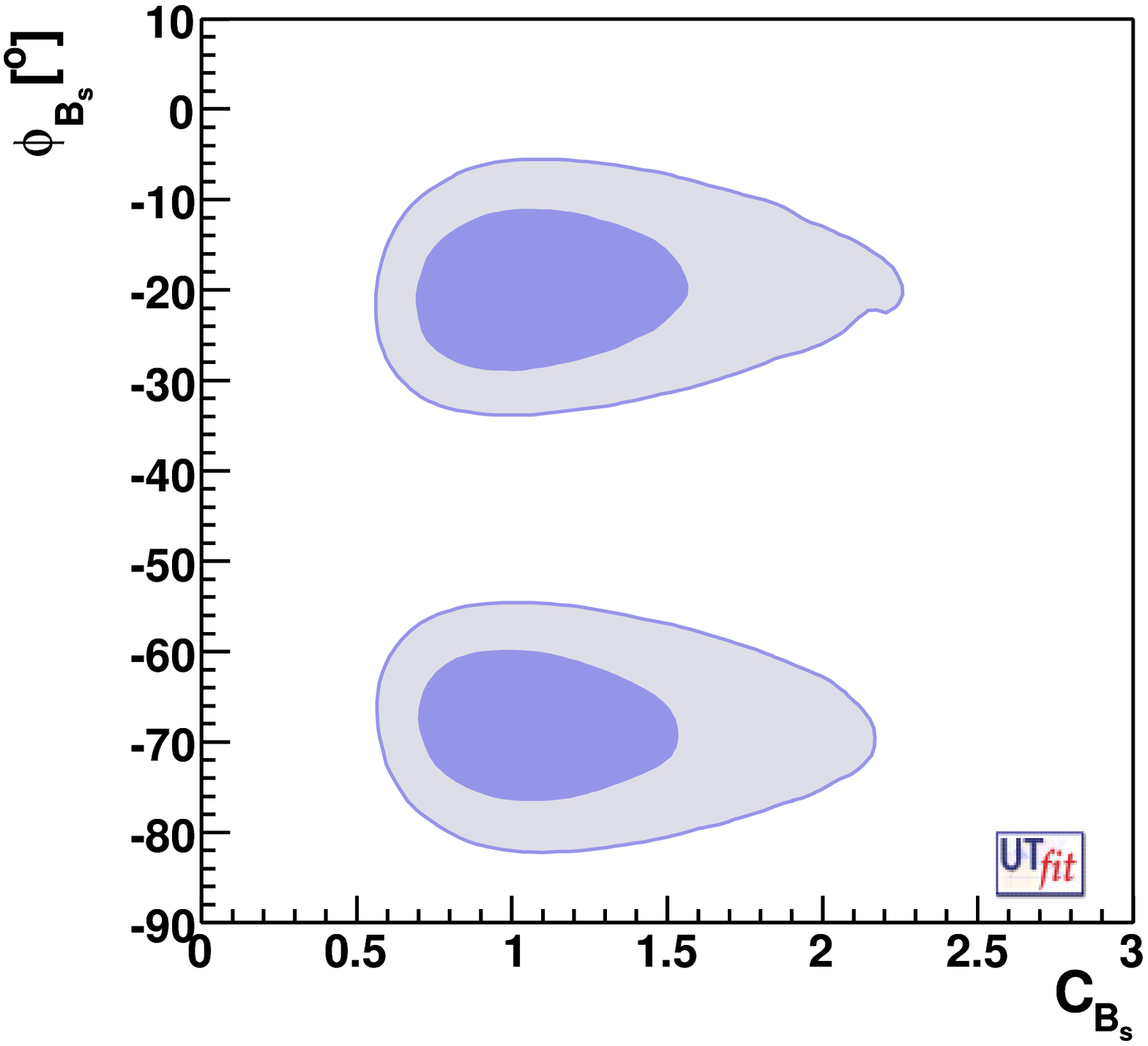}
\includegraphics[width=0.23\textwidth]{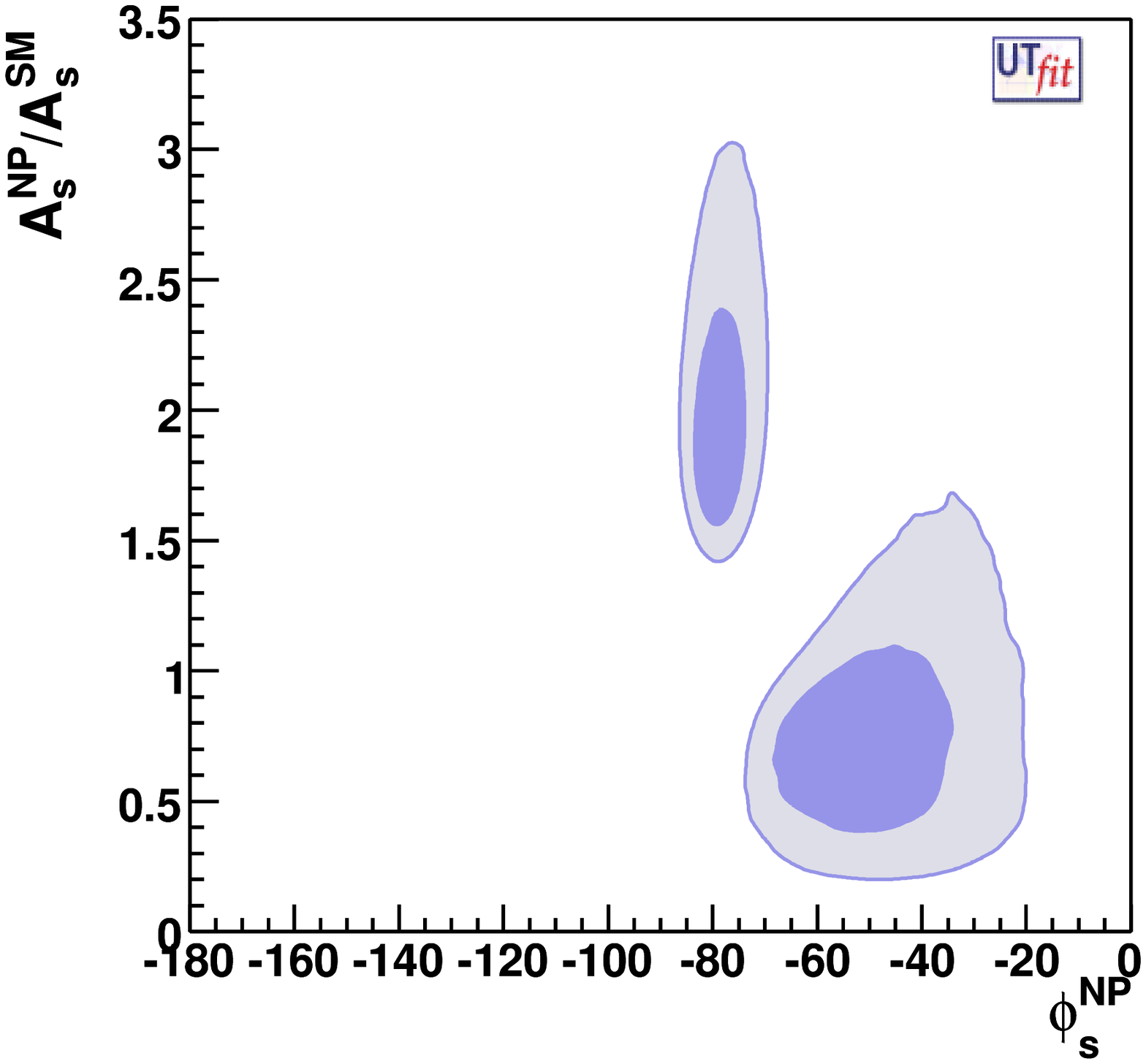}
\includegraphics[width=0.23\textwidth]{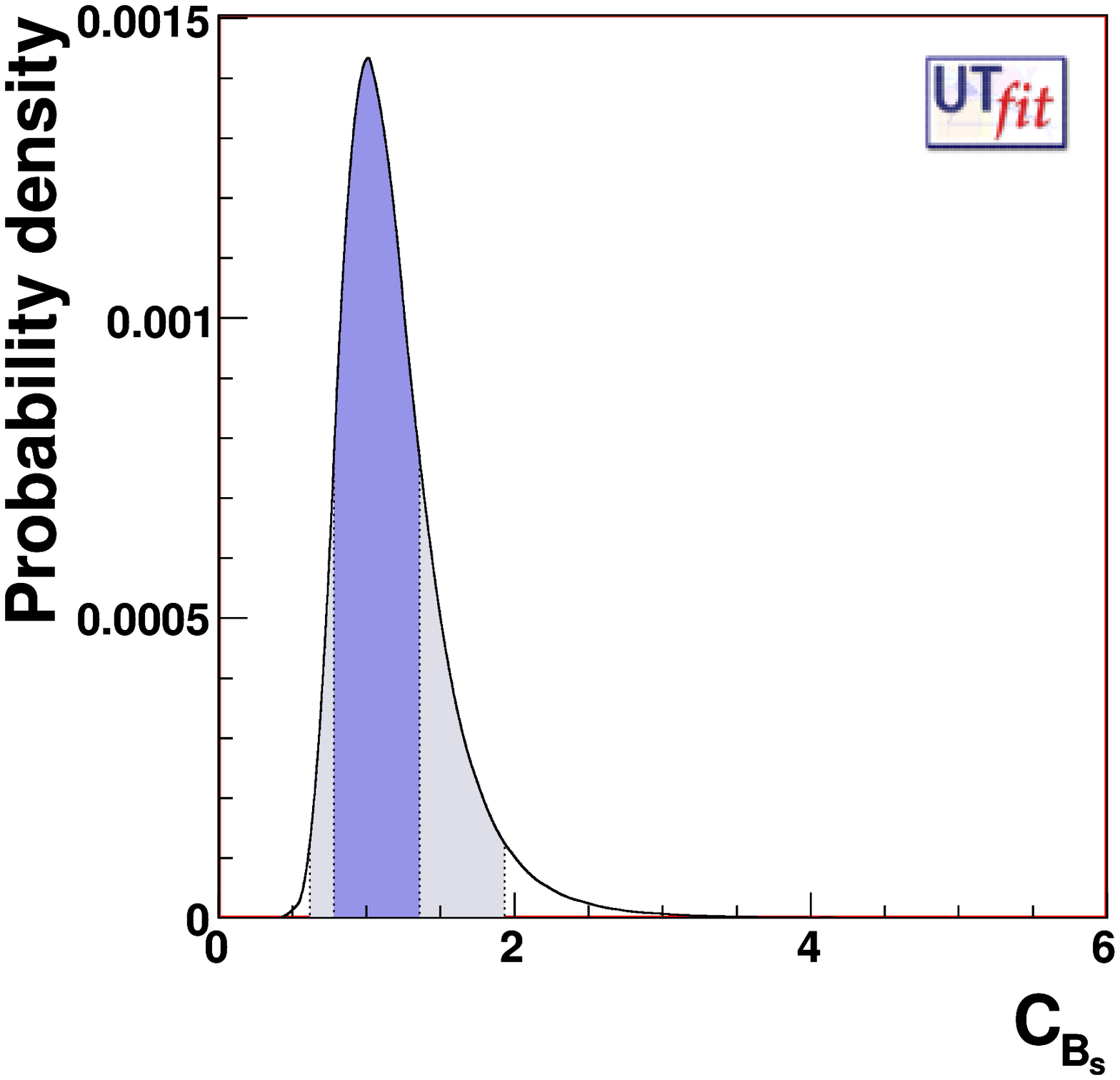}
\includegraphics[width=0.23\textwidth]{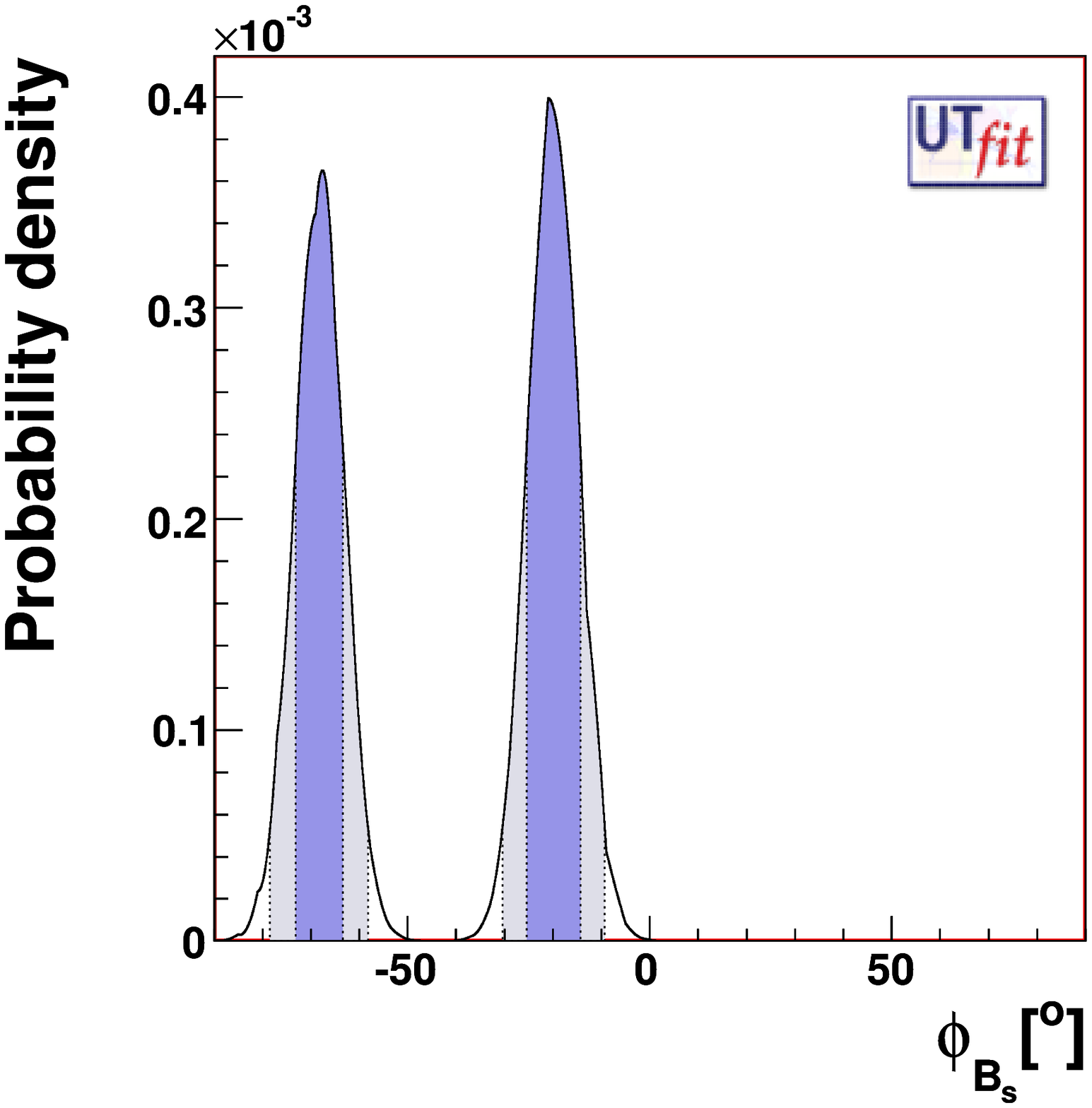} 
\includegraphics[width=0.23\textwidth]{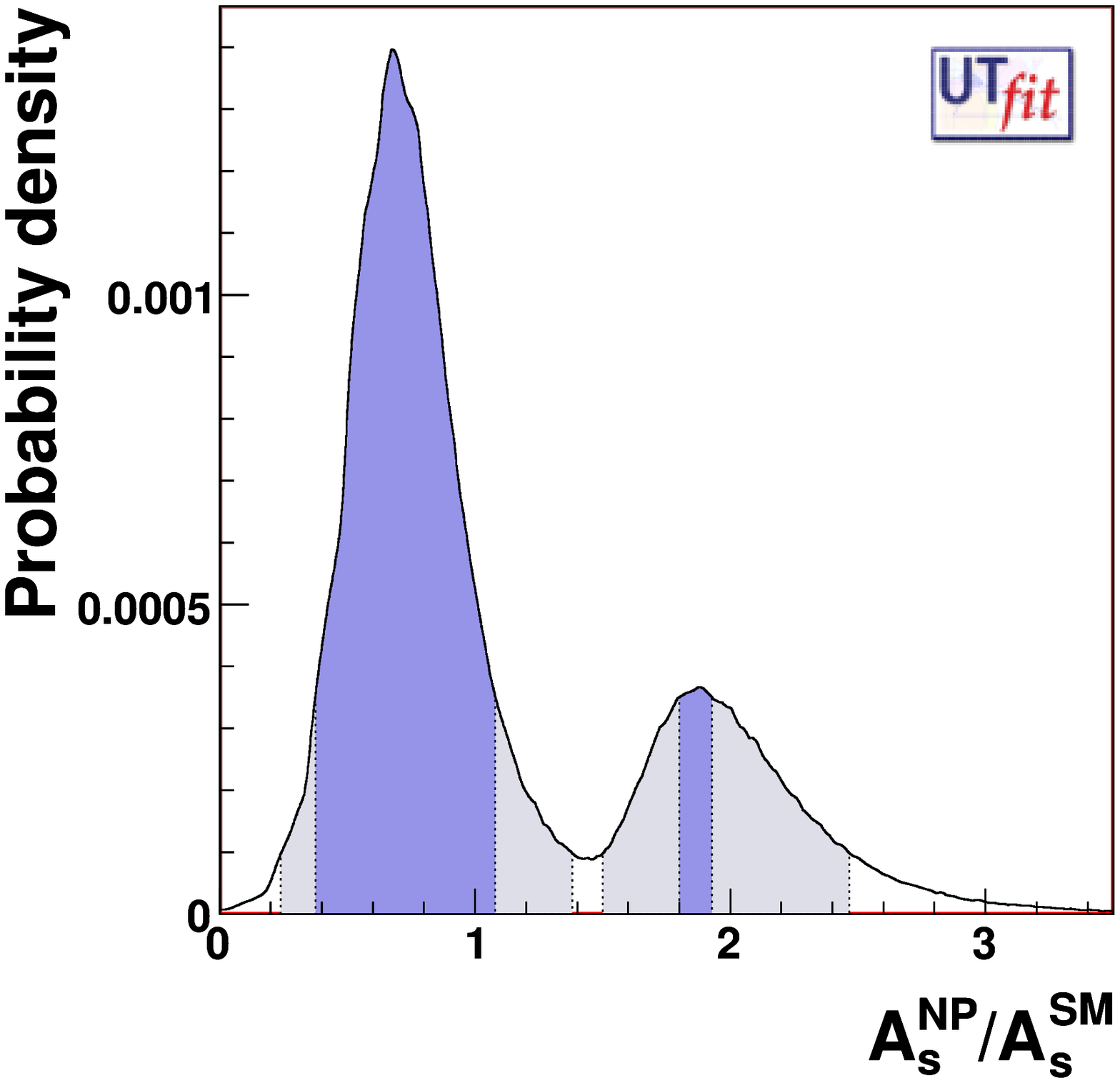}
\includegraphics[width=0.23\textwidth]{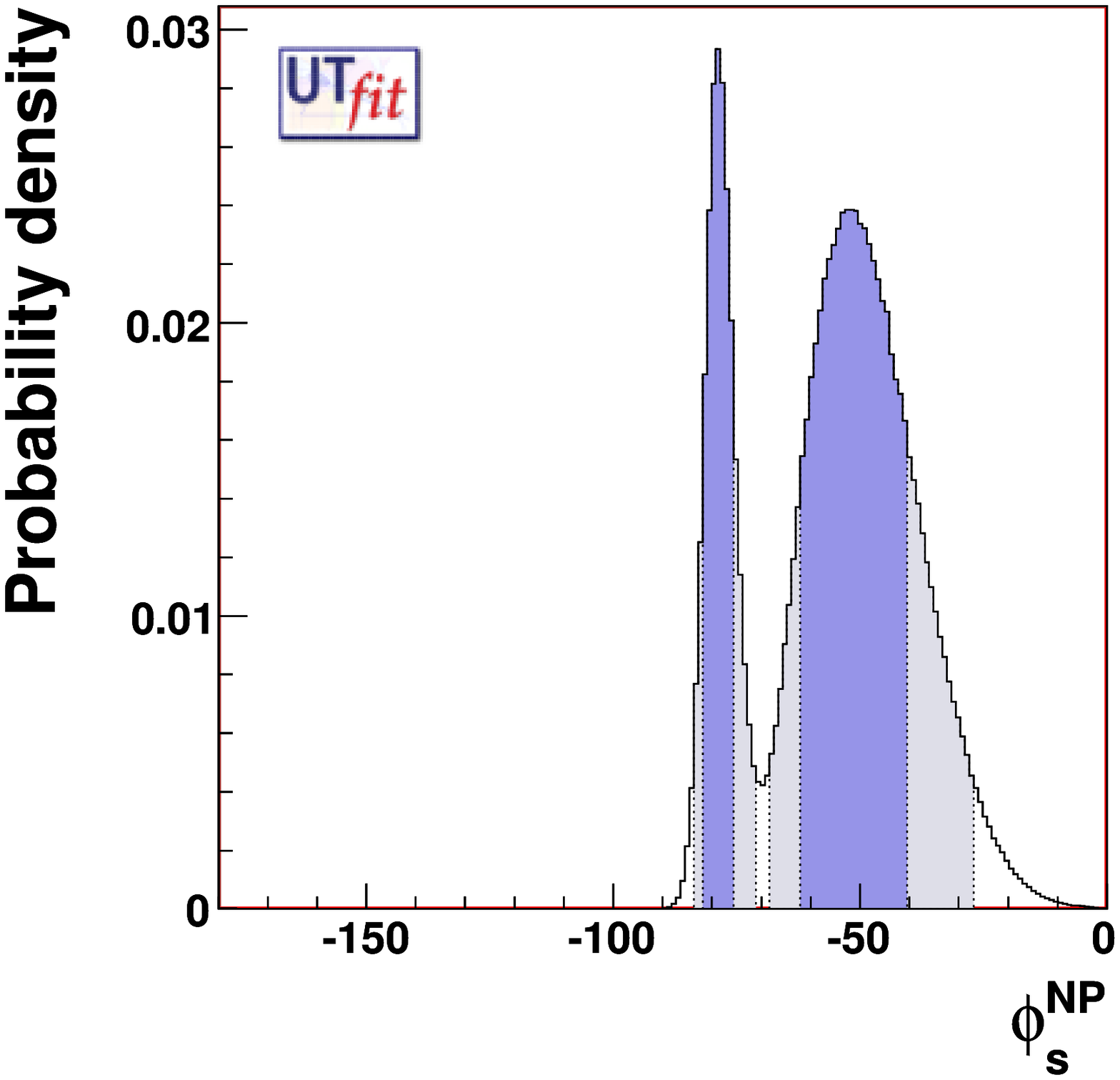}
\includegraphics[width=0.23\textwidth]{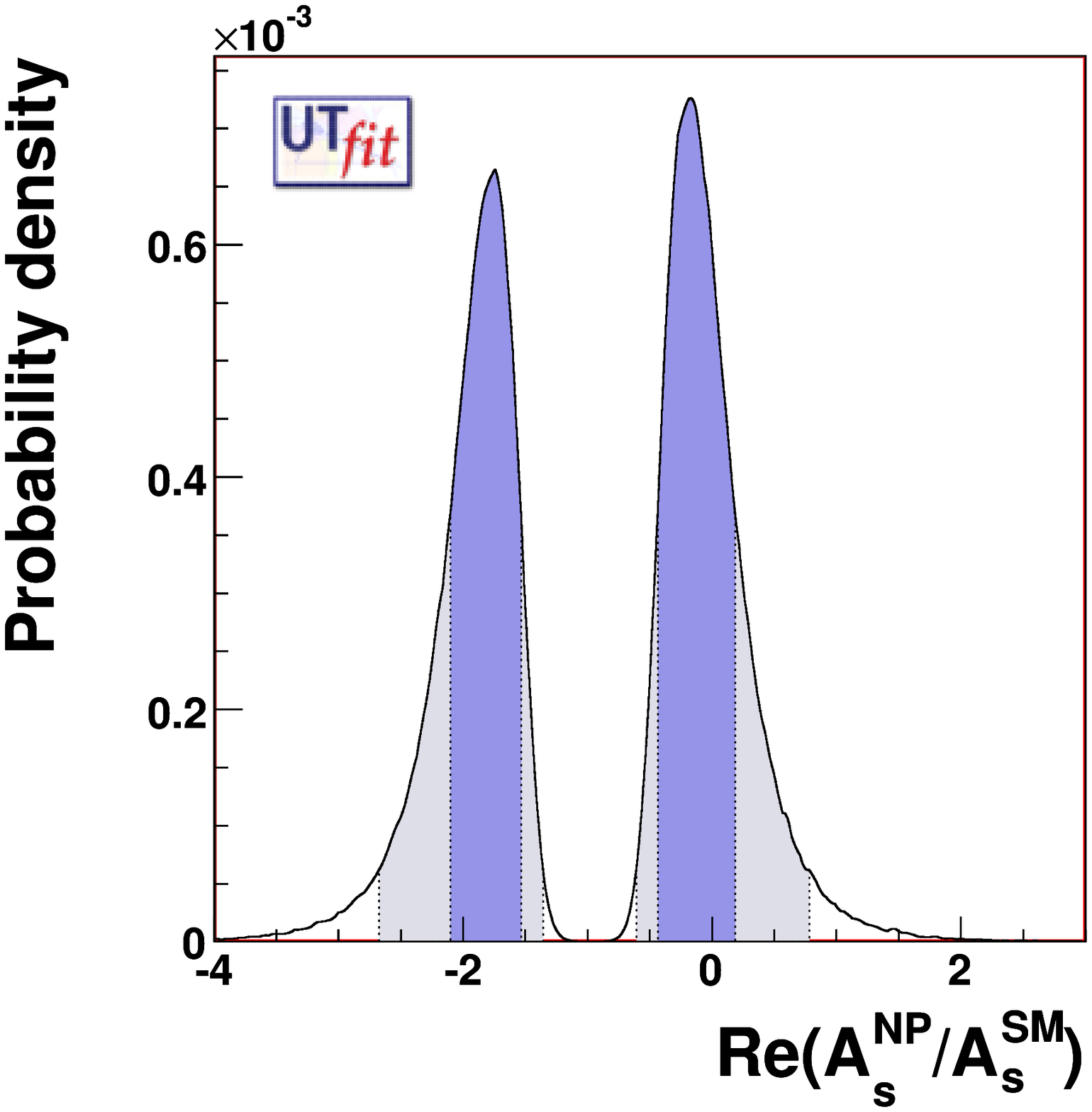}
\includegraphics[width=0.23\textwidth]{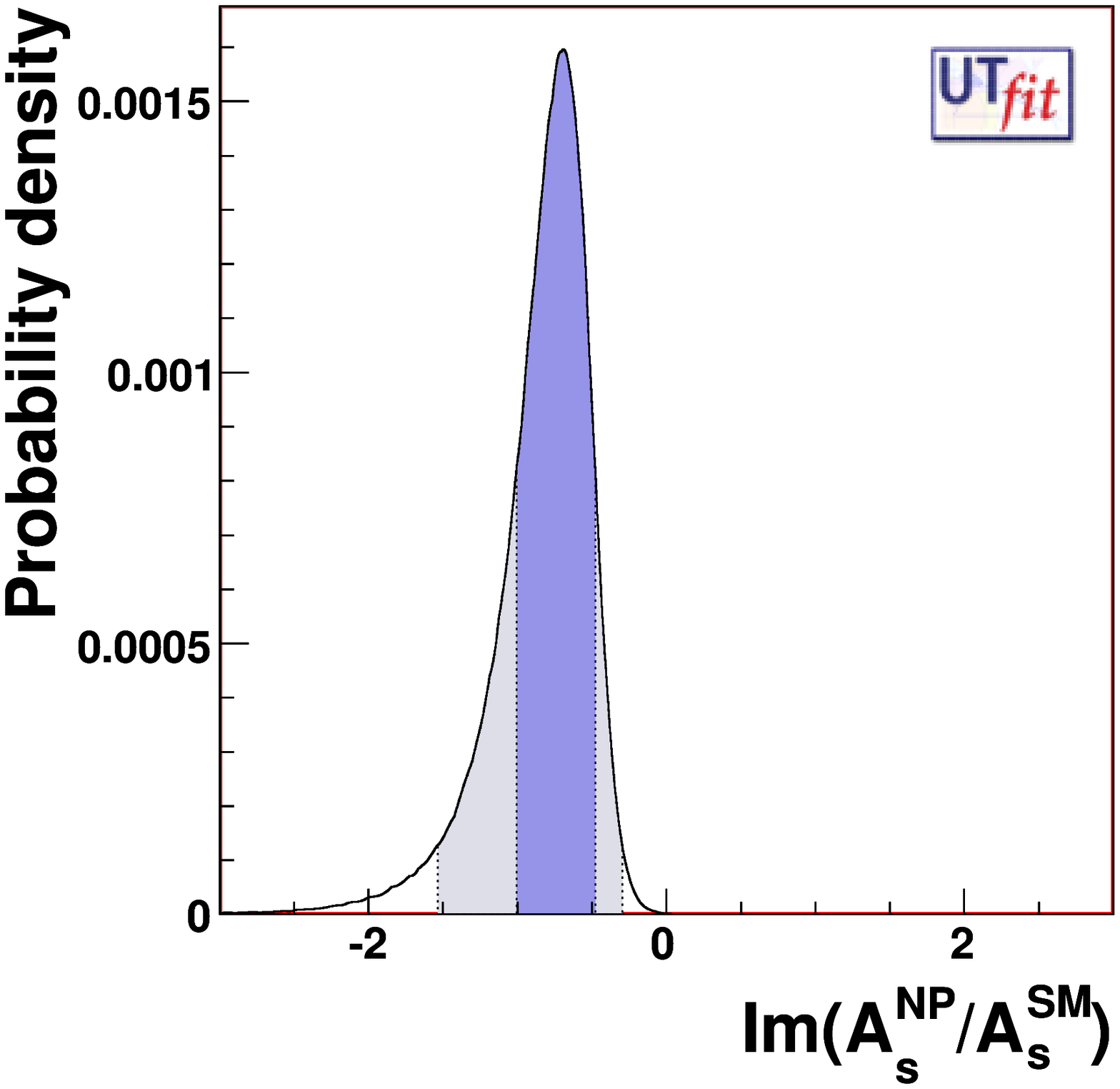}
\caption{%
  From left to right and from top to bottom, $68\%$ (dark) and  $95\%$
(light) probability regions in the
  $\phi_{B_s}$ -- $C_{B_s}$, $A^\mathrm{NP}_s/A^\mathrm{SM}_s$ --
  $\phi^\mathrm{NP}_s$ planes and p.d.f for  $C_{B_s}$, $\phi_{B_s}$,
  $A^\mathrm{NP}_s/A^\mathrm{SM}_s$, $\phi^\mathrm{NP}_s$, Re
  $A^\mathrm{NP}_s/A^\mathrm{SM}_s$, Im
$A^\mathrm{NP}_s/A^\mathrm{SM}_s$.}
\label{fig:NP}
\end{center}
\end{figure}

We use the following experimental input: the CDF measurement of
$\Delta m_s$~\cite{dmsCDF}, the semileptonic asymmetry in $B_s$ decays
$A_\mathrm{SL}^{s}$~\cite{ASLD0}, the dimuon charge asymmetry
$A_\mathrm{SL}^{\mu\mu}$ from D$\O$~\cite{ACHD0} and
CDF~\cite{ASLCDF}, the measurement of the $B_s$ lifetime from
flavour-specific final states~\cite{tauBsflavspec}, the
two-dimensional likelihood ratio for $\Delta \Gamma_s$ and
$\phi_s=2(\beta_s-\phi_{B_s})$ from the time-dependent tagged angular
analysis of $B_s\to J/\psi \phi$ decays by CDF~\cite{CDFTAGGED} and
the correlated constraints on $\Gamma_s$, $\Delta \Gamma_s$ and
$\phi_s$ from the same analysis performed by
D{\O}~\cite{D0TAGGED}. For the latter, since the complete likelihood
is not available yet, we start from the results of the $7$-variable
fit in the free-$\phi_s$ case from Table I of ref.~\cite{D0TAGGED}. We
implement the $7 \times 7$ correlation matrix and integrate over the
strong phases and decay amplitudes to obtain the reduced $3 \times 3$
correlation matrix used in our analysis. In the D{\O} analysis, the
twofold ambiguity inherent in the measurement ($\phi_s \to
\pi-\phi_s$, $\Delta \Gamma_s \to -\Delta \Gamma_s$, $\cos
\delta_{1,2} \to - \cos \delta_{1,2}$) for arbitrary strong phases was
removed using a value for $ \cos \delta_{1,2}$ derived from the BaBar
analysis of $B_d \to J/\Psi K^*$ using SU(3). However, the strong
phases in $B_d \to J/\Psi K^*$ and $B_s \to J/\Psi \phi$ cannot be
exactly related in the SU(3) limit due to the singlet component of
$\phi$. Although the sign of $\cos \delta_{1,2}$ obtained using SU(3)
is consistent with the factorization estimate, to be conservative we
reintroduce the ambiguity in the D{\O} measurement. To this end, we
take the errors quoted by D{\O} as Gaussian and duplicate the
likelihood at the point obtained by applying the discrete ambiguity.
Indeed, looking at Fig. 2 of ref.~\cite{D0TAGGED}, this seems a
reasonable procedure.  Hopefully D{\O} will present results without
assumptions on the strong phases in the future, allowing for a more
straightforward combination.  Finally, for the CKM parameters we
perform the UT analysis in the presence of arbitrary NP as described
in ref.~\cite{df2gen}, obtaining $\overline {\rho} = 0.140 \pm 0.046$,
$\overline {\eta} = 0.384 \pm 0.035$ and $\sin2\beta_s = 0.0409 \pm
0.0038$. The new input parameters used in our analysis are summarized
in Table \ref{tab:input}, all the others are given in
Ref.~\cite{df2gen}. The relevant NLO formulae for $\Delta \Gamma_s$
and for the semileptonic asymmetries in the presence of NP have been
already discussed in refs.~\cite{UTNP1,UTNP2,df2gen}.

\begin{table}[th]
\begin{center}
\begin{tabular}{@{}ccc}
Observable  & $68\%$ Prob. & $95\%$ Prob.  \\
\hline
$\phi_{B_s} [^\circ]$             & -19.9 $\pm$ 5.6 & [-30.45,-9.29]
\\
                                   & -68.2 $\pm$ 4.9 & [-78.45,-58.2]
\\
$C_{B_s}$                           & 1.07 $\pm$ 0.29 & [0.62,1.93] \\
\hline
$\phi^\mathrm{NP}_s  [^\circ]$      & -51 $\pm$ 11 & [-69,-27] \\
                                   & -79 $\pm$ 3 & [-84,-71] \\
$A^\mathrm{NP}_s/A^\mathrm{SM}_s$  & 0.73 $\pm$ 0.35 & [0.24,1.38] \\
                                   & 1.87 $\pm$ 0.06 & [1.50,2.47] \\
\hline
Im $A^\mathrm{NP}_s/A^\mathrm{SM}_s$ & -0.74 $\pm$ 0.26 &
[-1.54,-0.30] \\
Re $A^\mathrm{NP}_s/A^\mathrm{SM}_s$  & -0.13 $\pm$ 0.31 &
[-0.61,0.78] \\
                                   & -1.82 $\pm$ 0.28 & [-2.68,-1.36]
\\
\hline
$A_\mathrm{SL}^s \times 10^2$      & -0.34 $\pm$ 0.21 & [-0.75,0.03]\\
$A_\mathrm{SL}^{\mu\mu}\times 10^3$& -2.1 $\pm$ 1.0 & [-4.7,-0.3]\\
$\Delta \Gamma_s/\Gamma_s$         & 0.105 $\pm$ 0.049 & [0.02,0.20]
\\
                                   & -0.098 $\pm$ 0.044 &
[-0.19,-0.02]  \\
\hline
\end{tabular}
\end{center}
\caption {Fit results for NP parameters, semileptonic asymmetries and
width
  differences. Whenever present, we list the two solutions due to the
  ambiguity of the 
  measurements. The first line corresponds to the one closer to the
SM.}
\label{tab:results}
\end{table}

The results of our analysis are summarized in Table
\ref{tab:results}. We see that the phase $\phi_{B_s}$ deviates from
zero at $3.7\sigma$. We comment below on the stability of this
significance. In Fig.~\ref{fig:NP} we present the two-dimensional
$68\%$ and $95\%$ probability regions for the NP parameters $C_{B_s}$
and $\phi_{B_s}$, the corresponding regions for the parameters
$A^\mathrm{NP}_s/A^\mathrm{SM}_s$ and $\phi^\mathrm{NP}_s$, and the
one-dimensional distributions for NP parameters. Notice that the
ambiguity of the tagged analysis of $B_s \to J/\Psi \phi$ is slightly
broken by the presence of the CKM-subleading terms in the expression
of $\Gamma_{12}/M_{12}$ (see for example eq.~(5) of
ref.~\cite{UTNP2}).
The solution around $\phi_{B_s}
\sim -20^\circ$ corresponds to $\phi^\mathrm{NP}_s \sim -50^\circ$ and
$A^\mathrm{NP}_s/A^\mathrm{SM}_s \sim 75\%$.  The second solution is
much more distant from the SM and it requires a dominant NP
contribution ($A^\mathrm{NP}_s/A^\mathrm{SM}_s \sim 190\%$). In this
case the NP phase is thus very well determined. The strong phase
ambiguity affects the sign of $\cos \phi_s$ and thus Re
$A^\mathrm{NP}_s/A^\mathrm{SM}_s$, while Im
$A^\mathrm{NP}_s/A^\mathrm{SM}_s \sim -0.74$ in any case.

\begin{figure}[th]
\begin{center}
\includegraphics[width=0.23\textwidth]{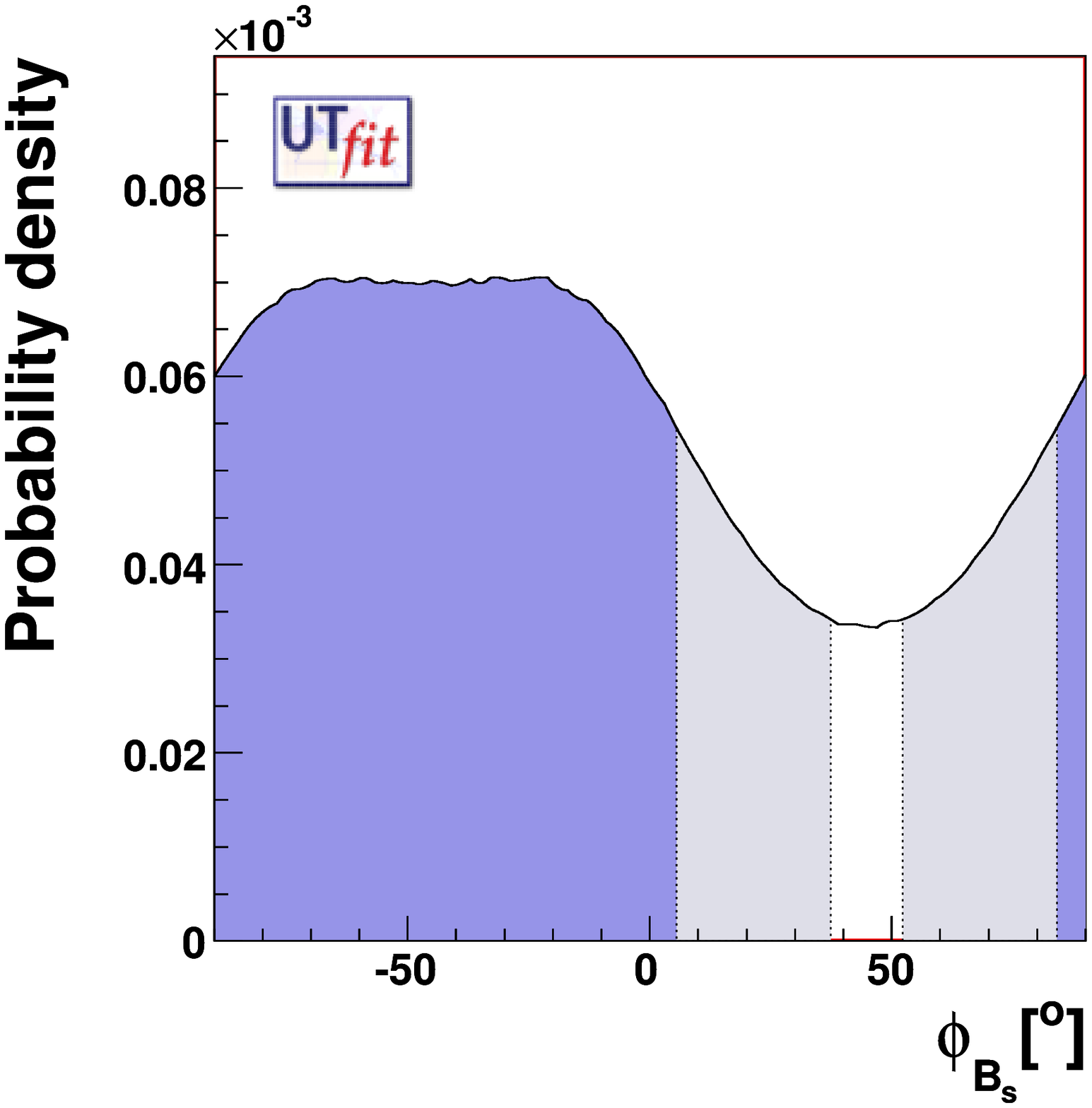}
\includegraphics[width=0.23\textwidth]{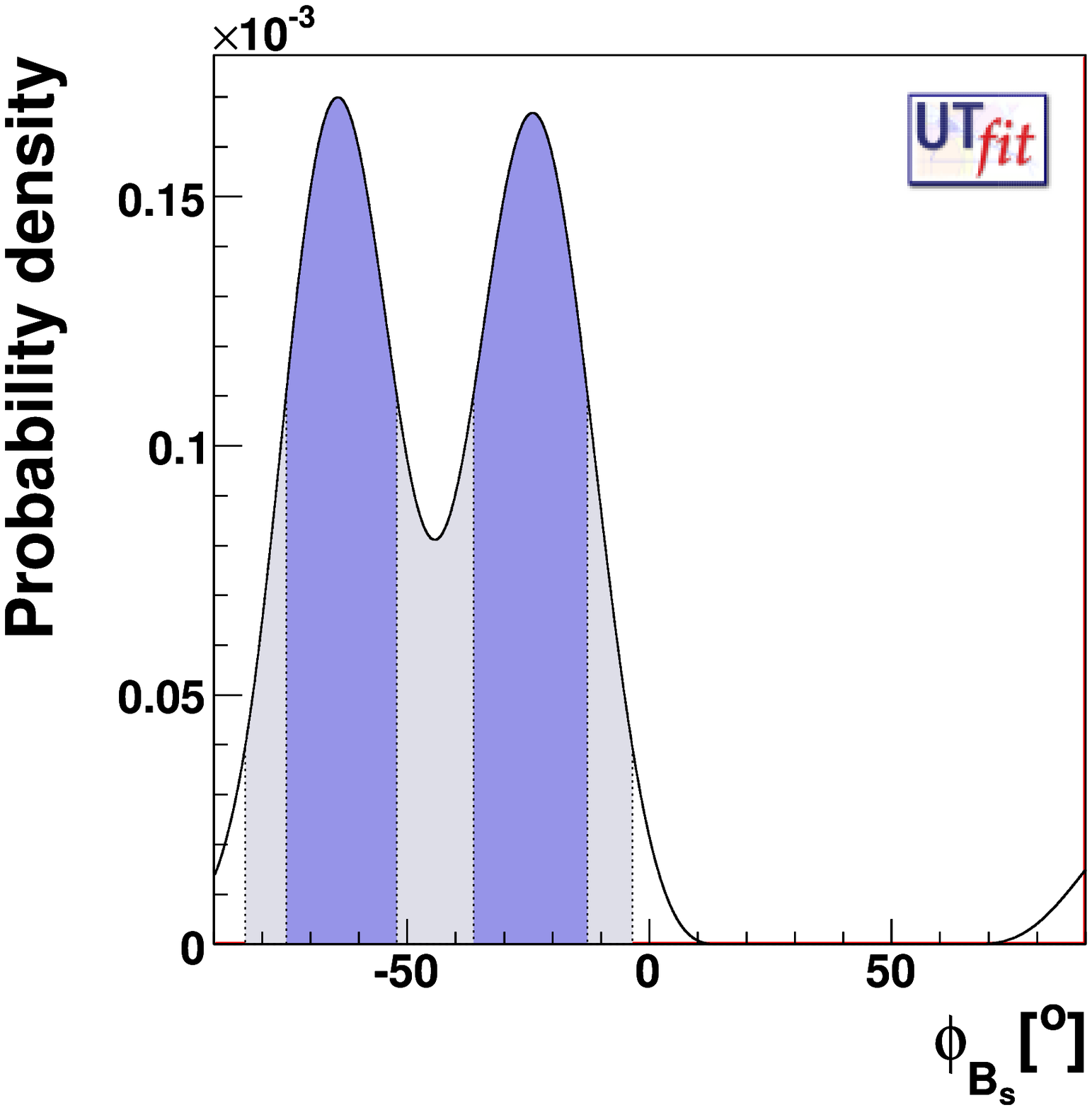}
\includegraphics[width=0.23\textwidth]{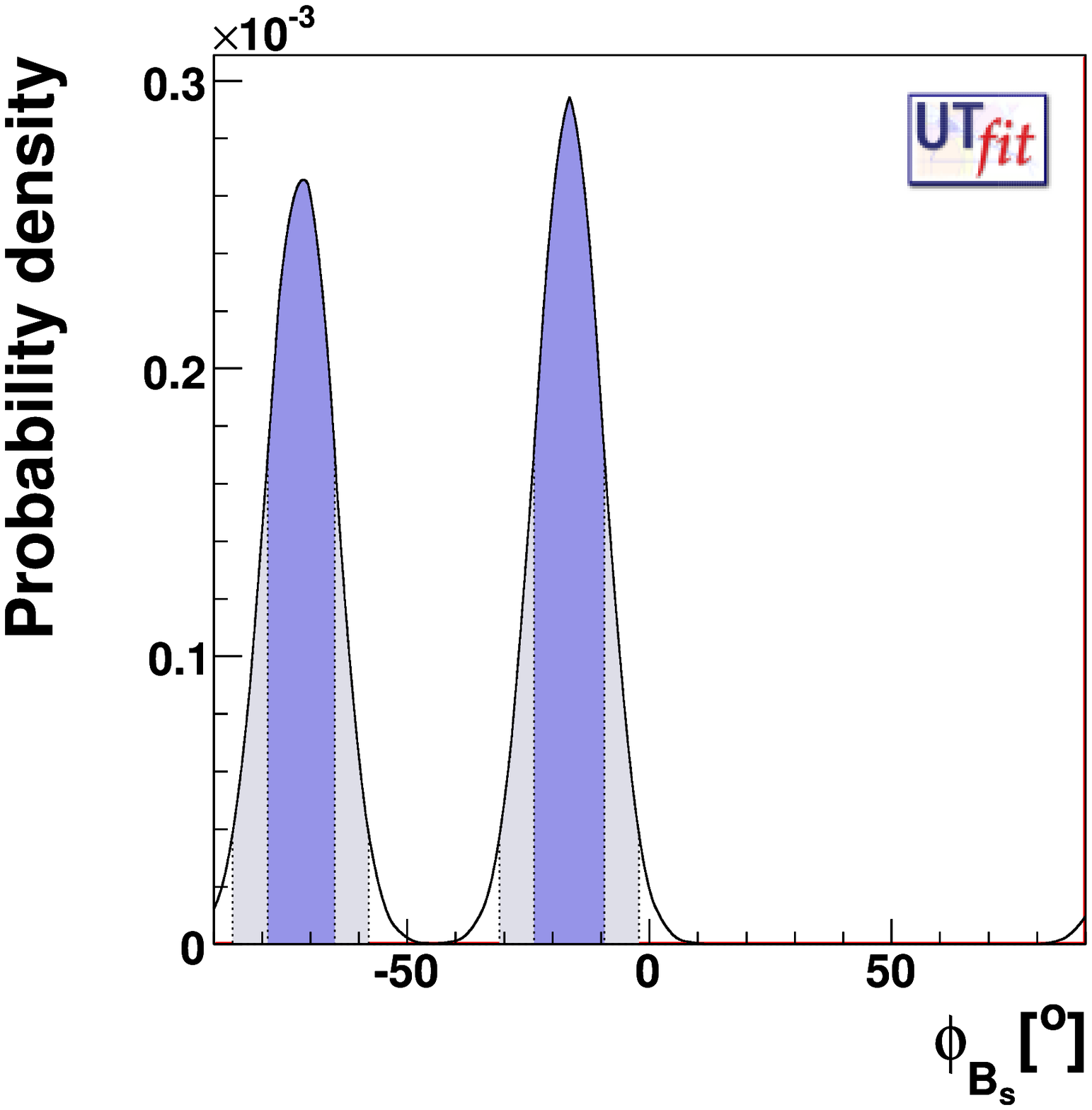}
\includegraphics[width=0.23\textwidth]{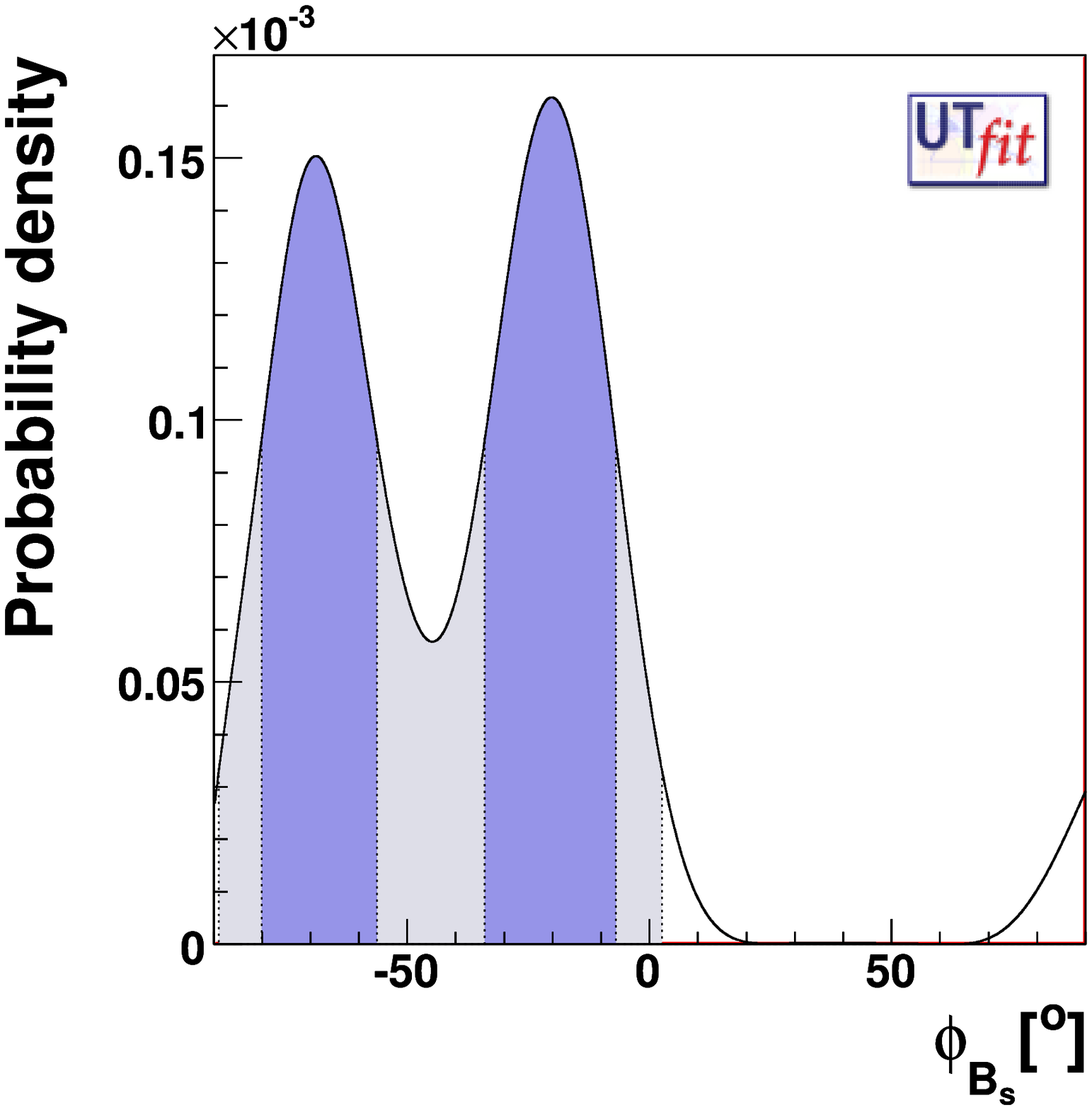}
\caption{%
  P.d.f. for $\phi_{B_s}$ without the tagged analysis of $B_s \to
  J/\Psi \phi$ (top left), including only the CDF analysis (top
  right), including only the D{\O} Gaussian analysis (bottom left),
  including only the D{\O} likelihood profiles (bottom right). We show
  $68\%$ (dark) and $95\%$ (light) probability regions.}
\label{fig:constraints}
\end{center}
\end{figure}

Before concluding, we comment on our treatment of the D{\O} result for
the tagged analysis and on the stability of the NP fit. Clearly, the
procedure to reintroduce the strong phase ambiguity in the D{\O}
result and to combine it with CDF is not unique given the available
information. In particular, the Gaussian assumption can be questioned,
given the likelihood profiles shown in Ref.~\cite{D0TAGGED}. Thus, we
have tested the significance of the NP signal against different
modeling of the probability density function (p.d.f.). First, we have
used the $90\%$ C.L. range for $\phi_s=[-0.06,1.20]^\circ$ given by
D{\O} to estimate the standard deviation, obtaining $\phi_s=(0.57 \pm
0.38)^\circ$ as input for our Gaussian analysis. This is conservative
since the likelihood has a visibly larger half-width on the side
opposite to the SM expectation (see Fig.~2 of Ref.~\cite{D0TAGGED}).
Second, we have implemented the likelihood profiles for $\phi_s$ and
$\Delta \Gamma_s$ given by D{\O}, discarding the correlations but
restoring the strong phase ambiguity.  The likelihood profiles include
the second minimum corresponding to $\phi_s \to \phi_s + \pi$, $\Delta
\Gamma \to - \Delta \Gamma$, which is disfavoured by the oscillating
terms present in the tagged analysis and is discarded in our Gaussian
analysis. Also this approach is conservative since each
one-dimensional profile likelihood is minimized with respect to the
other variables relevant for our analysis. It is remarkable that both
methods give a deviation of $\phi_{B_s}$ from zero of $3\, \sigma$
(the $3\, \sigma$ ranges for $\phi_{B_s}$ are $[-88,-48]^\circ \cup
[-41,0]^\circ$ and $[-88,0]^\circ$ for the two methods
respectively). We conclude that the combined analysis gives a stable
evidence for NP, although the precise number of standard deviations
depends on the procedure followed to combine presently available data.

To illustrate the impact of the experimental constraints, we show in
Fig.~\ref{fig:constraints} the p.d.f.\ for $\phi_{B_s}$ obtained
without the tagged analysis of $B_s \to J/\Psi \phi$ or including only
CDF or D{\O} results. Including only the CDF tagged analysis, we
obtain $\phi_{B_s} < 0$ at 97.7\% probability (2.3$\sigma$). For
D{\O}, we show results obtained with the Gaussian and likelihood
profile treatment of the errors. In the Gaussian case, the D{\O}
tagged analysis gives $\phi_{B_s} < 0$ at 98.0\% probability
(2.3$\sigma$), while using the likelihood profiles $\phi_{B_s} < 0$ at
92.8\% probability (1.8$\sigma$).  Finally, it is remarkable that the
different constraints in Fig.~\ref{fig:constraints} are all consistent
among themselves and with the combined result. We notice, however,
that the top-left plot is dominated by the measurement of
$A_\mathrm{SL}^{\mu\mu}$ while $A_\mathrm{SL}^s$ favours positive
$\phi_{B_s}$, although with a very low significance.  For
completeness, in Table \ref{tab:results} we also quote the fit results
for $A_\mathrm{SL}^s$, $A_\mathrm{SL}^{\mu\mu}$ and for $\Delta
\Gamma_s/\Gamma_s$.

In this Letter we have presented the combination of all available
constraints on the $B_s$ mixing amplitude leading to a first evidence
of NP contributions to the CP-violating phase. With the procedure we
followed to combine the available data, we obtain an evidence for NP
at more than $3\sigma$. To put this conclusion on firmer grounds, it
would be advisable to combine the likelihoods of the tagged $B_s \to
J/\Psi\phi$ angular analyses obtained without theoretical
assumptions. This should be feasible in the near future. We are eager
to see updated measurements using larger data sets from both the
Tevatron experiments in order to strengthen the present evidence,
waiting for the advent of LHCb for a high-precision measurement of the
NP phase.

It is remarkable that to explain the result obtained for $\phi_s$, new
sources of CP violation beyond the CKM phase are required, strongly
disfavouring the MFV hypothesis. These new phases will in general
produce correlated effects in $\Delta B=2$ processes and in $b\to s$
decays.  These correlations cannot be studied in a model-independent
way, but it will be interesting to analyse them in specific extensions
of the SM. In this respect, improving the results on CP violation in
$b\to s$ penguins at present and future experimental facilities is of
the utmost importance.

\section*{Note added}
During the review procedure of this Letter, results based on new data
were presented by the Tevatron experiments, as well as a combination of
Tevatron results on the tagged angular analysis of $B_s\to J/\psi\phi$.
However these updates are all unpublished. Furthermore, the likelihoods
required by our analysis are not publicly available except for the new
D{\O} analysis with no assumption on strong phases~\cite{d0new}.
For the sake of completeness, we quote $\phi_{B_s}=(-19\pm 8)^\circ
\cup (-69\pm 7)^\circ$ ($[-36,-5]^\circ \cup [-83,-54]^\circ$ at
$95\%$ probability),
obtained using this new likelihood for the D{\O} tagged angular analysis
of $B_s\to J/\psi\phi$. Clearly, we no longer need to manipulate the
D{\O} to remove the strong phase assumption and to account for the
non-Gaussian shape as described above. Remarkably, this updated result
is well compatible with the results of this Letter, confirming a
deviation from the SM at the level of $\sim 3\sigma$ ($99.6\%$ probability).
More recent experimental results seem to confirm the effect discussed in
this Letter. We will include them in future analyses as soon as
they become available.

We are much indebted to M.~Rescigno for triggering this analysis and
for improving it with several valuable suggestions.  We also thank
G.~Giurgiu, G.~Punzi and D.~Zieminska for their assistance with the
Tevatron experimental results. We acknowledge partial support from RTN
European contracts MRTN-CT-2006-035482 ``FLAVIAnet'' and
MRTN-CT-2006-035505 ``Heptools''. M.C.{} is associated to the
Dipartimento di Fisica, Universit\`a di Roma Tre. E.F.{} and L.S.{} are
associated to the Dipartimento di Fisica, Universit\`a di Roma ``La
Sapienza''.

\end{document}